\documentstyle[12pt]{article}
\newcommand{\be}{\begin{equation}}
\newcommand{\ee}{\end{equation}}
\newcommand{\bear}{\begin{eqnarray}}
\newcommand{\ear}{\end{eqnarray}}
\begin{document}
\begin{flushright}
HD-THEP-95-4\\
(revised)
\end{flushright}
\vspace{1cm}
\begin{center}
{\Large \bf Form Factor Relations for Heavy-to-Light Transitions$^*$}\\
\vspace{1.5cm}
Berthold Stech\\
\vspace{.5cm}
Institut f\"ur Theoretische Physik, Universit\"at Heidelberg$^+$,\\
 Philosophenweg 16, D-69120 Heidelberg\\
\vspace{1cm}

\noindent
{\bf Abstract:}
\end{center}
Assuming simple properties of the spectator particle in weak decays
the form factors of hadronic
current matrix elements are shown to be related
to a single universal function. The Isgur-Wise result for heavy-to-heavy
transitions follows as well as similar relations for heavy-to-light
decay processes. The approximation should hold for total energies
of the final particle large compared to the confinement scale.
A comparison with  experimentally determined
$D$-decay form factors and QCD sum rule results for $B$-decays
is very encouraging.
\vfill
\noindent$^*${\small Supported in part by the
Bundesministerium f\"ur Forschung und Technologie, Bonn}\\
\noindent$^+${\small E-mail: B.Stech@thphys.uni-heidelberg.de}

\newpage
The dynamical content of hadronic
current matrix elements is described by

Lorentz-invariant form factors. Knowledge of these
form factors is essential for the description of semileptonic
and nonleptonic weak decay processes and

in particular for the experimental determination of the fundamental

Kobayashi-Maskawa matrix elements. For transitions containing

one (infinitely) heavy
quark in the initial and another heavy quark in the final state

(heavy-to-heavy

transitions) the number of relevant form factors

is greatly reduced. For instance, in the limit $m_c\gg\bar\Lambda,
m_b\gg\bar\Lambda$

(with $\bar\Lambda\simeq m_D-m_c,\  m_B-m_b)$ the 6 form
factors describing
$\bar B\to D$ and $\bar B\to D^*$ vector and axial
vector matrix elements are

are all related to a single unknown form factor, the Isgur-Wise

function  \cite{1}. Even though $m_c$ is not really
large and leads to sizeable $1/m_c$ corrections the
Isgur-Wise relations
provide for a good starting point for more detailed
investigations based on
the heavy quark effective theory (HQET) \cite{2}.

For heavy to light transitions such as $B\to\pi$, $B\to \rho$,
on the other

hand, similar relations among the form factors cannot
be derived by using the
heavy quark limit. The heavy quark symmetries are
not applicable and the
number of independent form factors is not reduced.

Nevertheless, in the present note I will show that
interesting relations
among heavy-to-light form factors can be obtained, if use is made
of a constituent quark picture.

Let us consider the four momentum of a $B$-meson of mass $m_B$

and velocity $v^B$ and divide it into the momenta of the constituent
$b$-quark $p^B_b$ and the spectator $p^B_{sp}$
\bear\label{1}
&&P^B=p^B_b+p^B_{sp}\nonumber\\
&&p^B_b=\epsilon^B_b v^B+k^B,\quad p^B_{sp}=\epsilon^B_{sp}
v^B-k^B\nonumber\\
&&\epsilon^B_b+\epsilon^B_{sp}=m_B\ear
Here, $\epsilon_b^B$ and $\epsilon^B_{sp}$ denote the constituent
masses of $b$-quark and spectator within the $B$-meson, respectively.
The bound state dynamics is contained in the distribution function
for the off-shell momentum $k^B$, i.e. the wave function
of the $B$-meson. The four momentum of the final particle $F$

(with mass $m_F$ and velocity $v^F$) emitted in the weak process
is similarly decomposed. It contains the spectator particle plus
an $u$-quark in a $b\to u$ transition or a $c$-quark in
a $b\to c$ decay:
\bear\label{2}
&&P^F=p^F_{u,c}+p^F_{sp}\nonumber\\
&&p^F_{u,c}=\epsilon_{u,c}^Fv^F+k^F,\quad p^F_{sp}=
\epsilon^F_{sp}v^F-k^F
\nonumber\\
&&\epsilon^F_{u,c}+\epsilon^F_{sp}=m_F\ear
At this point I will make two dynamical assumptions:

i) In the rest system of a hadron the distribution of the components
of $k^\mu$ are strongly peaked with a width corresponding to the
confinement scale. (The values of the spectator masses
$\epsilon^{B,F}_{sp}\ll\epsilon^B_b$ are chosen such that the
peak is at $k^{B,F}_\mu=0$).
This assumption is plausible considering
the numerical value of the average
$b$-quark root mean square longitudinal
momentum $\sqrt{<p_z^2>}\approx 0.4$ GeV as obtained
from QCD sum rules \cite{R1}. It implies the dominance of
soft gluon effects over hard gluon emission before and

after the weak process.

ii) During the weak transition the spectator --- whatever it consists
of --- retains its momentum and spin. This requirement is
clearly satsified in any Fock space calculation of the
transition amplitude.

As a consequence of ii) the momenta $k^F$ and $k^B$ are correlated
in the weak process:
\be\label{3}
k^B-\epsilon^B_{sp}v^B=k^F-\epsilon^F_{sp}v^F\ee
While initial and final wave functions have their peaks at

$\bar k^B=\bar k^F=0$,
due to eq. (3) the integrand of the transition amplitude has
a maximum (with a width $\stackrel{\scriptstyle<}{\sim}
\bar\Lambda$) for values of $\bar k^B$ and

$\bar k^F$ different from zero.

Still, these values are such that --- in the rest system
of the $B$-meson --- both sides of eq. (3) stay of order
$\epsilon^{B,F}_{sp}\ll\epsilon^B_b$, even for the most
energetic transitions. To illustrate this, we consider, as
an example, a Gauss form of initial and final
wave functions. $\bar k_B$ and $\bar k_F$ are then determined
by the minimum of
\be\label{4}\alpha\left[2(k^F\cdot v^F)^2-(k^F)^2\right]
+2(k^B\cdot v^B)^2-(k^B)^2\ee
where $\alpha\approx 1$ denotes the ratio of the square of final

and initial particle radii. In the $B$-meson rest system
and for the spatial momentum of the final particle pointing in
$z$-direction, the result for $\bar k^B$ is\footnote{For simplicity,
we put $\alpha=1$. A change of this value or a different
choice $(>1)$ for the factors 2 in the expression (4) do not
invalidate the conclusions given below.}
\bear\label{5}
(\bar k^B)_0=\frac{1}{2}\epsilon^B_{sp}-\frac{1}{2}\frac{m_F}{E_F}
\epsilon^F_{sp},&&(\bar k^B)_z=-\frac{1}{2}\frac{P_F}{E_F}
\epsilon^B_{sp}\nonumber\\
(\bar k^B)_\perp&=&0.\ear
$E_F$ and $P_F$ denote energy and momentum of the
final particle. Thus,
the relevant $b$-quark space momenta active in the transition and
the $b$-quark energy variations are small
even in a transition with large energy
release! We can neglect $\bar k^B$ compared to the $b$-quark mass.

Using eq. (2,3), one can now estimate the momentum range of the

generated $u$- or $c$-quark and finds that --- in the $B$-meson
rest frame --- it is peaked around
\bear\label{6}
(\bar p_{u,c})_0&=&E_F(1+O(\epsilon_{sp}^{F,B}/E_F))\nonumber\\
(\bar p_{u,c})_z&=&P_F(1+O(\epsilon_{sp}^{F,B}/E_F))\ear
Thus, the $u$- or $c$-quark momenta effectively determining
the weak transition amplitude lie close to the 4-momentum of the

final particle --- apart from corrections of order $\epsilon_{sp}
/E_F$.

{}From this result it is easy to obtain form factor relations in
the limit where $(\epsilon_{sp}^{F,B})^2$ and the average

transverse quark momentum squared are small compared to $E^2_F$:
The transition matrix element of the weak current is simply

proportional to the $c$-number matrix element $T^\mu$
\be\label{7}
T^\mu=\left(\bar u_{u,c}^{s'}(\vec P_F,m_{u,c})\gamma^\mu(1-\gamma_5)
u^s_b(\vec 0,m_b)\right)L_{s',s}.\ee
Here $m_{u,c}$ and $m_b$ are current masses of the $u$ or $c$-quark and
the $b$-quark, respectively. The energy $\bar p^0_{u,c}\simeq\sqrt
{\vec P_F^2+\vec k^2_\perp+m^2_{u,c}}$ must be identified with $E_F$.
The $L_{s's}$ are the elements of a $2\times 2$
spin matrix with $L=1\!{\rm l}$ for

$B$-decays to a pseudoscalar state, e.g. the $\pi$-meson, and
$L=\sigma\cdot\vec e$ for $B$-decays to a vector particle polarized

in $\vec e$ direction. The form (7) satisfies the requirement that the
spin components of the spectator particle remain unaffected in
the decay process.

A comparison of (7) with the conventional form factor decomposition
\cite{3} of the current matrix element gives for a transition to
a pseudoscalar particle (the $\pi$ or the $D$, for instance)
\bear\label{8}
&&F_1(q^2,m_F)=\left(1+\frac{m_{u,c}}{m_B}\right)R^B_{u,c}
(q^2,m_F)\nonumber\\
&&F_0(q^2,m_F)=\left(1+\frac{m_{u,c}}{m_B}-\frac{q^2}{m_B^2-m^2_F}
(1-\frac{m_{u,c}}{m_B})\right)R^B_{u,c}(q^2,m_F).\ear
$R^B_{u,c}(q^2,m_F)$ is an unknown universal function depending not
only on $q^2$ but also on $m_F$ and the flavor of the
outgoing quarks (and on $m_B$).

For transitions to a vector particle
(the $\rho$ or $D^*$ for instance) I find with $E_F=\frac{1}
{2m_B}(m_B^2+m_F^2-q^2)$
\bear\label{9}
&&V(q^2,m_F)=\frac{m_B+m_F}{m_B}R^B_{u,c}(q^2,m_F)\nonumber\\
&&A_1(q^2,m_F)=2\frac{m_{u,c}+E_F}{m_B+m_F}R^B_{u,c}
(q^2,m_F)\nonumber\\
&&A_2(q^2,m_F)=\frac{m_B+m_F}{m_B}\ \frac{E_F-\frac{m^2_F}{m_B}+
m_{u,c}\left(1+\frac{m_F}{m_B}\right)}{m_F+E_F}
R^B_{u,c}(q^2,m_F)\nonumber\\
&&A_0(q^2,m_F)=\frac{m_{u,c}+m_B}{m_B}R^B_{u,c}(q^2,m_F).\ear

In the limit of large $m_c$  one can use the approximation
$m_D=m_{D^*}=m_c$ i.e. apply the spin symmetry of HQET valid for
heavy-to-heavy transitions.
Eqs. (8) and (9) then give
\bear\label{10}
F^{HH}_1(q^{{2}})&=&\left(1+\frac{m_D}{m_B}\right)
R^B_c(q^2,m_F)\nonumber\\
F^{HH}_0(q^2)&=&A^{HH}_1(q^2)=\left(1-q^2/(m_B+m_D)^2\right)
F^{HH}_1(q^2)\nonumber\\
V^{HH}(q^2)&=&A^{HH}_2(q^2)=A^{HH}_0(q^2)=F^{HH}_1(q^2).\ear
Thus the well-known heavy-to-heavy form factor relations \cite{4}
based on the heavy quark limit \cite{1} are contained in (8) and (9).

For $b\to u$ transitions we can set $m_u=0$ and find
\bear\label{11}
&&F_1^{HL}(q^2,m_F)=R^B_u(q^2,m_F)\nonumber\\
&&F_0^{HL}(q^2,m_F)=\left(1-\frac{q^2}{m_B^2-m^2_F}\right)
F_1^{HL}(q^2,m_F)
\nonumber\\
&&V^{HL}(q^2,m_F)=\left(1+\frac{m_F}{m_B}\right)
F^{HL}_1(q^2,m_F)\nonumber\\
&&A^{HL}_1(q^2,m_F)=\frac{1+\frac{m^2_F}{m^2_B}}
{1+\frac{m_F}{m_B}}\left
(1-\frac{q^2}{m^2_B+m^2_F}\right)F^{HL}_1(q^2,m_F)\nonumber\\
&&A^{HL}_2(q^2,m_F)=\left(1+\frac{m_F}{m_B}\right)\left(
1-\frac{2m_F/(m_B+m_F)}{1-q^2/(m_B+m_F)^2}\right)
F^{HL}_1(q^2,m_F)\nonumber\\
&&A^{HL}_0(q^2,m_F)=F^{HL}_1(q^2,m_F).\ear
Remarkably, the heavy-to-light form factor relations are
not very different from
to the heavy-to-heavy form factor relations (10).
But, of course, here the dependence of $F_1$ on $m_F$ has

to be taken into account due to the lack of spin symmetry in the
final state.

Of particular interest is the fact that the longitudinal form factor
$F_0$ and the transverse form factor $A_1$ again behave  differently
from the remaining form factors. This result is strongly supported
by a recent detailed QCD sum rule calculation of $B\to\rho$
form factors by P. Ball \cite{5}. She found a strong difference
between the $q^2$-dependence of $A_1$ and the other $B\to \rho$
form factors. Similar results have been obtained in ref.
\cite{R2}, \cite{R3}. Ref. \cite{R3} indicates that also the
form factor $A_0$ satisfies eq. (11).

The differential branching ratio for a semileptonic
decay to a vector particle

using Eq. (9) is now (in a more general notation
and neglecting the lepton mass):
\bear\label{12}
\frac{d\ BR(q^2)}{dq^2}&=&\frac{G^2_F}{192\pi^3}\frac{\lambda(q^2)}
{m_I^5}\tau_I|V_{fi}|^2\left(R^I_f(q^2,m_F)\right)^2\cdot \left(
S_T(q^2)+S_L(q^2)\right)\nonumber\\
\lambda(q^2)&=&[(m_I+m_F)^2-q^2]^{1/2}\cdot[(m_I-m_F)^2-q^2]^{1/2}.
\ear
Here $V_{fi}$ denotes the relevant Cabibbo-Kobayashi-Maskawa
matrix element,
$G_F$ the Fermi constant and $\tau_I$ the lifetime of the initial
pseudoscalar meson. $S_T(q^2)$ and $S_L(q^2)$ are the transverse
and longitudinal polarization contributions, respectively:
\bear\label{13}
S_T(q^2)&=&2q^2\left[\lambda^2(q^2)+4m^2_I(m_f+E_F(q^2))^2\right]
\nonumber\\
S_L(q^2)&=&\left[(m^2_I-m_F^2)(m_I+m_f)-q^2(m_I-m_f)\right]^2.\ear
$m_f$ denotes the \underline{current} mass of the emitted quark
active in the process. The differential branching ratio for a

decay to a pseudoscalar particle is obtained from (12) by
replacing the sum $S_T+S_L$ by

\be\label{14}
S_P(q^2)=\lambda^2(q^2)m^2_I\left(1+\frac{m_f}{m_I}\right)^2.\ee
Eqs. (8-14)
are expected to hold to a good approximation for $E_F\gg \epsilon
_{sp}\approx0.35$ GeV.
The only unknown is the function $R^I_f(q^2,m_F)$.

The assumptions leading to the result (8-14)
are rather general. It should hold or approximately hold in all

quark model calculations which treat the spectator the same way as
done here and use the relativistic Dirac-spinor structure.
An interesting publication by the Orsay group \cite{6} deals with
an explicit semi-relativistic wave function model  which  gives a

decreasing $q^2$-behaviour of $A_1(q^2)/V(q^2)$ and reproduces
for large $m_F$
the Isgur-Wise relations. Their formulae differ,
however, for heavy-to-light transitions from the ones found here
since in their model the light quark is not treated in a fully
relativistic manner. Another model worth mentioning here is the
one by Faustov and Galkin \cite{R4}.

One may be hesitant to apply the result (8,9) for $D$-decays
(replacing $m_B$ by $m_D$ and $m_u$ by $m_s$) because of the
relatively low energies of the final particles involved. Let us
nevertheless try it. Using $m_s=m_{D_s}-m_{D^+}=0.10$ GeV
one obtains for $D\to K^*$ transitions at
$q^2=0$ the form factor
values shown in Table I.

\bigskip
\begin{center}
{\bf Table I:} $D\to K^*$ form factors at $q^2=0$\\
\bigskip
\begin{tabular}{l|c|l}
& \ theory (Eq. (9))& \ experiment \cite{7}\\
\hline
$V(0)\ $&$\ 1.00\cdot\delta$&$\ 1.16\pm0.16$\\
$A_1(0)\ $&$\ 0.61\cdot\delta$&$\ 0.61\pm0.05$\\
$A_2(0)\ $&$\ 0.42\cdot\delta$&$\ 0.45\pm0.09$\\
\end{tabular}
\end{center}
\bigskip

For the ratio of form factors the agreement
with experiment\footnote{The data at $q^2=0$

are extracted from integrated rates assuming single pole

formulae. Thus, they are  not completely free of theoretical

uncertainties.} is surprisingly
good. Moreover, I will show below that $\delta$ defined by
$\delta=R_s^D(0,m_{K^*})/0.67$ can be estimated and turns out
to be very close to one.

For $B$-decays to light particles there are not yet

experimental data available to test Eq. (11). One can,
however, compare form factor ratios from (11) with the explicit
QCD sum rule calculations of ref. \cite{5} and ref. \cite{6a}.
Table II shows as

representative examples the $B\to\rho$ transition form factors
at $q^2=0$ and at $q^2=8$ GeV$^2$.
Noticeably, there is agreement with the QCD sum rule result.
In particular, in all three calculations the ratio $A_1/V$ falls
off with $q^2$.

\bigskip
\begin{center}
{\bf Table II:} $B\to \rho$ form factors at $q^2=0$ and $q^2=8
\ {\rm GeV}^2$\\
\bigskip
\begin{tabular}{l|c|c|c}
& \ theory (Eq. (11)) &\  ref. \cite{5} &\ ref. \cite{6a}\\
\hline
$A_1/V|_{q^2=0}$&$\ 0.78$&$\ 0.83\pm0.32$&$\ 0.86\pm0.23$\\
$A_2/V|_{q^2=0}$&$\ 0.75$&$\ 0.67\pm0.40$&\\
$A_1/V|_{q^2=8\ {\rm GeV}^2}\ $&$\ 0.56$&$\ 0.50\pm0.19$&$
\ 0.60\pm0.19$\\
$A_2/V|_{q^2=8\ {\rm GeV}^2}\ $&$\ 0.67$&$\ 0.66\pm0.39$&
\end{tabular}
\end{center}
\bigskip

Encouraged by the above success one can go further
and can try to relate heavy-to-heavy with heavy-to-light form
factors. This requires, however, a new dynamical assumption
referring to the formation of the final particle, e. g. a fixing of

initial and final wave functions. As a first attempt let us
simply ignore the different structure and radii of the final
hadrons of a given spin (say a $\rho$ and
a $D^*$). Under this condition $R^B_{u,c}(q^2,m_F)$ depends only on
the velocity of the outgoing hadron apart from the explicit
dependence on the quarks contained in the Dirac spinors of Eq. (7).
Since $R$ does scale with $m_B^{1/2}$, Eq. (7) together with

dimensional arguments allow to write $R$ in the form
\bear\label{15}
&&R^B_{u,c}(q^2,m_F)=\frac{1}{2}\left(\frac{m_B}{E_F+m_{u,c}}
\right)
^{1/2}(1+y)^{1/2}\ \xi(y)\nonumber\\
&&y=\frac{E_F}{m_F}=v_F\cdot v_B=\frac{m^2_B+m^2_F-q^2}
{2m_Bm_F}.\ear
Here $\xi(y)$ is now a universal function solely dependent
on $y$, i.e. the same for $\bar B\to D^*,\bar B\to\rho$
and $D\to K^*$
transitions. The prefactors in (15) have been chosen in such
a way that $\xi(y)$ is just the Isgur-Wise function as can be
seen by comparing (15) with (10) and using the conventional
definition of this function for heavy-to-heavy
transitions.
Clearly, a direct practical use of comparing a $b\to c$ with a
$b\to u$ or $c\to s$ transition can only be
made if the values of $y$
considered belong to the physical region of both processes.
Moreover, good results can only be expected if $E_F\gg
\epsilon_{sp}^{B,F}$ holds for the heavy-to-light transition.

As a simple test for the applicability of (15) one can take
the numerical value of the Isgur-Wise
function for the $\bar B\to D^*$ transition at a given value of $y$

in order to obtain the form factors for $\bar B\to\rho$ decays at
the corresponding $q^2$ value.
For $y=1.5$ (i.e. $q^2=0$ for the $\bar B\to D^*$ decay) the
corresponding momentum transfer in the $\bar B\to\rho$ transition
is $q^2=16.3\ {\rm GeV}^2$. Taking $V_{cb}\cdot\xi(1.5)=0.023
\pm0.002$ \cite{8}
and $V_{cb}=0.040\pm0.003$ \cite{9},
Eq. (15) gives $R^B_u(16.3,m_\rho)=0.97\pm0.11$.
{}From (11) one then gets the $B\to\rho$ form factor values shown in

Table III.
Remarkably, the theoretical numbers are in agreement with
the values obtained from the plots in ref. \cite{5} and ref.

\cite{6a}.
\newpage
\begin{center}
{\bf Table III:} $B\to \rho$ form factors at $q^2=16.3$ GeV$^2$\\
\bigskip
\begin{tabular}{l|c|c|c}
& \ theory (Eq. (11))\  & \ ref. \cite{5}\ & \ ref. \cite{6a} \\
\hline
$V^{B\to\rho}(16.3)\ $&$\ 1.11\pm0.13$&$\ 0.96\pm0.32$&$\ 1.55\pm0.50$\\
$A_1^{B\to\rho}(16.3)\ $&$\ 0.37\pm0.04$&$\ 0.30\pm0.06$&$0.50\pm0.05$\\
$A_2^{B\to\rho}(16.3)\ $&$\ 0.60\pm0.07$&$\ 0.52\pm0.26$&\\
\end{tabular}
\end{center}
\bigskip

By using the equivalent of (15) for $D$-decays
\be\label{16}
R^D_s(q^2,m_F)=\frac{1}{2}\left(\frac{m_D}{E_F+m_s}\right)^{1/2}
(1+y)^{1/2}\ \xi(y)\ee
it is also possible to get from $\xi(y)$ information
on $D\to K^*$ transitions.
Choosing $y=1.28$ which corresponds to $q^2=0$
for the $D\to K^*$ transition
and taking $V_{cb}\cdot\xi(1.28)=0.029\pm0.003$ \cite{8},
Eq. (16) leads
to $R^D_s(0,m_{K^*})=0.67\pm0.08$. Thus, the quantity
$\delta$ defined earlier is obtained to be $\simeq1$ giving
close agreement between theory and experiment in

$D$-decays\footnote{A more direct way to obtain $R^B_u$ and $R^D_s$
from (15,16) is to extract $R^B_c(q^2,m_{D^*})$ from the

measured differential branching ratio \cite{9a} according to Eq. (12).}.

Heavy-to-light
current matrix elements are also needed in the calculation of
nonleptonic and Penguin-induced matrix elements \cite{3}.
Of recent interest \cite{6, 10}
are the decays $B\to K^{(*)}J/\psi$ which are given --- in factorization
approximation --- by the $B\to K^{(*)}$ form factors \cite{3, 6, 10}.
For the calculation of the polarization of
 $K^*$ one needs only
the ratio of form factors (at $q^2=m^2_{J/\psi})$.
It can be
directly  obtained from (13). For the longitudinal
polarization $\rho_L$
one gets $\rho_L=0.41$ not in accord with the most recent value
$\rho_L=0.66\pm0.1\pm0.1$ \cite{11} or the even larger values of

previous Argus and Cleo results \cite{12}.
I do not consider the result for the longitudinal polarization as an

argument against (9). Factorization is an
approximate concept \cite{3},
and the longitudinal polarization involving the interference of
$S$ and $D$ waves
is particular sensitive to final state interactions.
The small class II transitions can always get
corrections from the stronger class I
transitions to (virtual) intermediate  $D^{(*)}\bar D
_s^{(*)}$-like states turning into $K^*J/\psi$.

For Penguin-induced processes like $\bar B\to K^*"\gamma"$

Isgur and Wise \cite{12a} derived relations between
the corresponding
form factors and the form factors of semi-leptonic decays.
In particular, the magnetic moment form factor

$F^{magn.mom.}(q^2)$ --- as defined  for instance
in ref. \cite{6a} --- is related to the form factors
$A_1(q^2)$ and $V(q^2)$:
\be\label{17}
F^{magn.mom.}(q^2)=\frac{m_I+m_F}{2m_I}A_1(q^2)+
\frac{V(q^2)}{2m_I}\frac{q^2+m^2_I-m^2_F}
{m_I+m_F}\ee
Originally derived for $q^2\approx q^2_{max}$ this equation
is also supposed to hold for small $q^2$ since --- according
to Burdman and Donoghue \cite{R5} --- hard perturbative
contributions may be neglected. QCD sum rule calculations \cite{6a}
indeed show that eq. (17) holds to good accuracy for all momentum
transfers. The present investigation supports this result. Moreover,
for sufficiently large $E_F$ one gets using (8,9)
\be\label{18}
F^{magn.mom.}(q^2,m_F)=F_1(q^2,m_F)=\left(1+\frac{m_f}{m_I}
\right)R^I_f(q^2,m_F)\ee
thus providing for a simple connection between the branching ratios of
radiative and semileptonic decays.

The formulae given in this paper give a handle on
heavy-to-light matrix elements. Eqs. (8-14) combined
with constraints from dispersion theory \cite{13}

will be useful for the determination of the
Kobyashi-Maskawa matrix element $V_{ub}$.

In addition --- but with less rigor --- one can make use
of the universality property expressed in (15, 16):
Taking a dispersion theoretic formula for the vector
form factor $R(q^2)$ in (12) and fitting the

corresponding parameters to a single decay mode many predictions can
be made. However, more work is necessary to get a precise control of

the theoretical errors.

\noindent{\bf Acknowledgement:} It is a pleasure to thank
Volker Rieckert for valuable discussions.

\end{document}